\documentclass{ws-mpla}

\begin{document}

\renewcommand{\draftnote}{}
\renewcommand{\trimmarks}{}
\renewcommand{\catchline}{}

\markboth{E. A. Matute} {Presymmetry Beyond the Standard Model}

\catchline{}{}{}{}{}

\title{PRESYMMETRY BEYOND THE STANDARD MODEL}

\author{\footnotesize ERNESTO A. MATUTE}

\address{Departamento de F\'{\i}sica, Universidad de Santiago de
Chile, Usach,\\ Casilla 307 -- Correo 2, Santiago, Chile\\
ernesto.matute@usach.cl}

\maketitle

\pub{}{}

\begin{abstract}
We go beyond the Standard Model guided by presymmetry, the
discrete electroweak quark--lepton symmetry hidden by topological
effects which explain quark fractional charges as in condensed
matter physics. We show that partners of the particles of the
Standard Model and the discrete symmetry associated with this
partnership appear as manifestations of a residual presymmetry in
the sense of Ekstein and its extension from matter to forces. This
duplication of the spectrum of the Standard Model keeps spin and
comes nondegenerated about the TeV scale.

\end{abstract}

\ccode{PACS Nos.: 11.30.Hv, 11.30.Ly, 12.10.Dm, 12.60.-i}

\section{Introduction}

In the phenomenological successful Standard Model (SM), the number
of fermion generations is not fixed by any symmetry principle. The
famous question ``who order that'' of I. I. Rabi when the muon was
identified in the late '40s of the last century, has been
generalized to why are there three families, but remains
unanswered at the level of the SM. Moreover, constraints from high
precision experiments do not prohibit new sequential or
nonsequential families, but instead provide only restrictions on
their mass spectrum. The resulting masses for the new sequential
leptons should make easy their detection with neutrino pair
productions being the interesting signals. Nonsequential quarks
and leptons, partners of the SM ones, should be produced in pairs
if they are protected by some new discrete symmetry. These partner
particles, although more difficult to understand, are favored by
the existence of dark matter in the universe. However, with so
many quark--lepton family replicas it is conceivable that the
symmetries ordering the known fermion generations, on the one
hand, and the lightest new partner particles, on the other hand,
be related to each other in a unified description. Ultimately,
they would be part of the same family replication problem.

Partners of the SM particles have been suggested by various new
physics models motivated by different puzzling aspects of the SM.
One of these is the hierarchy problem,\cite{tH} i.e. the disparity
between the energy scale where electroweak symmetry breaking takes
place and the scale of new physics if the SM is viewed as an
effective theory with a cutoff that can be as low as a few TeV.
This discrepancy manifests itself through quadratic divergences on
the cutoff affecting the SM Higgs mass, required to be of a few
hundred GeV by electroweak precision measurements. The most
popular approaches to solve this naturalness problem are directed
by the supersymmetric models,\cite{SS1,SS2} in which fundamental
light Higgs bosons are maintained; the little Higgs
models,\cite{LH1,LH2} in which the light Higgs boson also appears
elementary but is identified as a pseudo-Goldstone boson in a
scenario where, however, the problem is only pushed to higher
scales; and the extra dimension models,\cite{KK} in which a
fundamental Planck scale close to the electroweak scale is
advanced, so that the ultimate ultraviolet cutoff is around the
TeV scale, protecting the Higgs mass from divergent radiative
corrections. To evade all constraints from electroweak precision
data, these proposals require the existence of heavy partners of
the SM particles at the Terascale and an associated extra $Z_2$
symmetry. This discrete symmetry is $R$-parity in supersymmetric
models,\cite{RP1,RP2} $T$-parity in little Higgs
models,\cite{TP1}$^{\mbox{--}}$\cite{TP3} and $KK$-parity in the
so-called universal extra dimension models\cite{KKP} which,
however, do not directly address the fine-tuning issue.
Nevertheless, it should be stressed that the hierarchy problem is
not indeed a trouble with the SM itself; it may not exist if there
is no new physics between the electroweak scale and the Planck
scale.\cite{Shapo} On the other hand, the idea of unification of
the SM gauge couplings is degraded in little Higgs models and
extra dimension models.

Another puzzling feature of the SM is the left--right asymmetry of
its weak interactions. Mirror matter models address this problem
by introducing $P$-parity and mirror partners for all SM
particles,\cite{PP} although this parity symmetry is violated in
each electroweak sector and renormalizable gauge-invariant
interactions between standard and mirror particles are possible
mainly through mixing terms involving the standard and mirror
Higgs fields. The physical Higgs boson can then decay into mirror
particles, i.e. invisibly, through all possible mechanisms of the
Higgs boson production itself, giving distinctive signatures of
the mirror world that can be tested.

There is still no experimental confirmation of any of these views
on nature, whose symmetries do not relate the partner particles
with the fermion family problem either. Hence, with the emerging
CERN--LHC era, it is important to explore any other well-motivated
scenario that gives rise to observable partners associated with an
underlying discrete symmetry which be somehow connected with the
origin of the triplication of fermion families. In this letter we
report achievements from models which address in the first place
the question of the quark--lepton symmetry, exhibited plainly in
the electroweak gauge sector of the SM when Dirac neutrinos are
included.

The discrete quark--lepton symmetry has recently been extended
from the weak to electromagnetic interactions by considering
topological effects as in condensed matter physics to account for
quark fractional charges.\cite{EAM1}$^{\mbox{--}}$\cite{EAM22} The
quark--lepton charge relations are explained by adding to the SM
with Dirac neutrinos the new hidden quark (lepton) states of
prequarks (preleptons) with integer (fractional) charges as in
leptons (quarks). By exploiting the discrete prequark--lepton
(quark--prelepton) charge symmetry, the so-called
presymmetry,\footnote{Here we clarify the concept of (charge)
presymmetry introduced within the framework of quarks and leptons
along the idea of presymmetry of Ekstein\cite{Ekstein1,Ekstein2}
inasmuch as we have, as shown in Sec.~2, residual (charge)
symmetry transformations in spite of the existence of
symmetry-breaking dynamical effects.} several other riddles of the
SM have been understood, such as the charge quantization,
confinement of fractional charges and triplication of
quark--lepton families, relating the number of generations with
the number of colors. The gauge anomalies associated with
prequarks (preleptons) lead to the appearance of a topological
charge, which induces the fractional charges and the physical
quark (lepton) states.\cite{EAM1} The universality of weak
interactions of quarks and leptons also holds at the level of
prequarks and preleptons, with weak topological charges being
generated upon ones and the others.

Electroweak presymmetry is hidden at the level of standard quarks
and leptons. Due to its topological character, however,
presymmetry is independent of the energy scale and therefore
underlies any new physics beyond the SM with Dirac neutrinos,
which usually invokes new matter and interactions. In this letter
we address ourselves to the possibility that traces of this
symmetry may show up at low energies from a symmetric expansion of
the SM gauge symmetry $\mbox{SU(3)}_{q} \times
\mbox{SU(2)}_{q\ell} \times \mbox{U(1)}_{Y}$. Specifically, we
consider the well-motivated symmetric model $\mbox{SU(3)}_{q}
\times \mbox{SU(3)}_{\tilde{q}} \times
\mbox{SU(2)}_{q\tilde{\ell}} \times \mbox{SU(2)}_{\tilde{q}\ell}
\times \mbox{U(1)}_{Y} \times \tilde{P}$ of standard and exotic
quarks and leptons of the same 1/2 spin,\cite{EAM3} where the
exotic partners are denoted by tildes and $\tilde{P}$ is a $Z_2$
discrete symmetry of the full Lagrangian under the exchange of the
new particles and the SM particles that constrains the bare color
as well as weak coupling constants to be equal.

Our aim in this letter is to show that the exotic partners and the
associated \linebreak $\tilde{P}$-symmetry, named exotic symmetry,
appear as manifestations of a remaining presymmetry in the sense
of Ekstein\cite{Ekstein1,Ekstein2} and its extension from matter
to forces. This will mean to add significant new understanding to
the physics of both presymmetry as presented in
Ref.~\refcite{EAM1} at the SM level and the exotic symmetry in the
exotic doubling of the SM proposed in Ref.~\refcite{EAM3}. In the
scenario of the above expanded gauge symmetry, exotic symmetry
involves new physics that can be close to the electroweak scale.
In fact, it has been shown that at the level of normalized quarks
and leptons, all the features of the SM are retained even if the
new neutral and charged weak bosons were relatively light. These
nonstandard bosons have the signature of exotic symmetry but they
do not exhibit the universality of the interactions of the
standard ones, so that a lower bound for their masses can be
set.\cite{EAM3} Also, the fermion partners fit the electroweak
data with mass below 1 TeV and are stable if the fermion numbers
are separately conserved. From the viewpoint of the hierarchy
problem mentioned above, the duplication of the SM with Dirac
neutrinos under the exotic symmetry remains natural at the
electroweak scale, deferring the fine-tuning problem of any other
gauge symmetry breaking at higher energies.

Our prime motivation to prompting new physics beyond the SM with
Dirac neutrinos by duplicating gauge groups with quark and lepton
type of families, as done in Ref.~\refcite{EAM3}, is indeed to
generate a residual presymmetry in the sense of Ekstein, so
avoiding the use against the model of the often-cited principle
known as Occam's razor\footnote{For a recent use of the Occam's
razor principle, see Ref.~\refcite{Occam}.}: ``Entities should not
be multiplied unnecessarily.'' In fact, if no extension of the SM
with Dirac neutrinos is considered, presymmetry remains as a
hidden feature of the SM with no direct implications to be
observed. Therefore, by Occam's razor it would have to be
eliminated from the SM, expecting that its indirect successful
implications\cite{EAM1} would be explained differently by some
other physics beyond the SM. In particular, the quark--lepton
charge symmetries presented in Refs.
\refcite{EAM1}--\refcite{EAM22} in support of presymmetry should
be taken as accidental and not real, which is difficult to accept.
Hence, doubling of the SM particles with a residual presymmetry
becomes a presymmetry requirement.

A second motivation for this doubling of the SM is to expand
presymmetry from matter to forces. Since presymmetry seems to be a
relevant hidden symmetry of nature and \emph{transverse to
everything}, it is feasible that this discrete symmetry partially
or completely extends to the forces of the SM, so that symmetric
fermions interact with symmetric gauge bosons and the
corresponding gauge symmetry is doubled. Also, it is conceivable
that the replication of fermionic families be accompanied of a
replication of bosonic force carriers, with a common underlying
symmetry ordering their occurrence.

A third motivation is to extend the discrete symmetry from the
electroweak to the strong sector, i.e. to have presymmetry for the
full Lagrangian of fundamental interactions, therefore acquiring
more significance with a strong influence on the course of the new
physics beyond the SM. Furthermore, from the point of view of the
presumptive existence of new generations, presymmetry requires
that the new quark families be nonsequential, which demands a
duplication of the color gauge group SU(3). This suggests in turn
a duplication of the electroweak group SU(2)$\times$U(1). The
extra quark and lepton partners avoid in an obvious way any
anomaly problem in the duplication of the gauge groups. Besides,
if new quark--lepton families are found, the topological formalism
to explain fractional charges also applies to quark partners, so
that the symmetry associated with their existence has to be
connected with presymmetry.

A fourth motivation comes from the fact that though standard
quarks and \linebreak leptons are subject to universality in the
Lagrangian of weak interactions, their partners should not because
these are not sequential fermions. It is quite possible that weak
universality is only a low-energy property of standard quarks and
leptons. This can be realized by duplicating the weak gauge
group.\cite{EAM3}

All of the above provides reasons to expect that at least a piece
of the new physics that can be explored around the TeV scale is a
manifestation of presymmetry. In the context of the extended gauge
model taken up in this letter, this is effected through the
relation between presymmetry and the exotic symmetry. On the other
hand, the single doubling of the list of the SM particles
anticipated by presymmetry does not contravene other models that
also introduce partner particles. It does not spoil, for instance,
that suggested by supersymmetry. Moreover, these doublings of the
particle spectrum may be compatible. Their motivations are after
all quite different: supersymmetry attacks a mass problem, whereas
presymmetry addresses a charge question.

The letter is organized as follows. The presymmetric expansion of
the SM in the scenario of hidden prequarks and leptons is
presented in Sec.~2 and that in terms of quarks and hidden
preleptons is given in Sec.~3, showing that the exotic partners
and the associated exotic symmetry appear as manifestations of a
residual presymmetry and its extension from matter to forces.
Phenomenological implications for searches of new physics about
the TeV scale are discussed in Sec.~4. Conclusions are drawn in
Sec.~5.

\section{Presymmetric Extension of the SM with Prequark Partners}

We start by describing the fermionic sector of the model with
hidden integer-charged prequarks underlying fractionally-charged
quarks. It is an enlargement to the extended gauge group of the
topological approach to charge structure of quarks developed at
the SM level.\cite{EAM21,EAM22}

The spectrum of physical fermions includes quarks, leptons with
Dirac neutrinos, and their exotic partners. Their assignments
under the extended gauge group $\mbox{SU(3)}_{q} \times
\mbox{SU(3)}_{\tilde{q}} \times \mbox{SU(2)}_{q\tilde{\ell}}
\times \mbox{SU(2)}_{\tilde{q}\ell} \times \mbox{U(1)}_{Y}$ are
\begin{eqnarray}
\begin{array}{rll}
& \displaystyle q^{i}_{nL} \sim \left( 3, 1, 2, 1, \frac{1}{3}
\right), & \qquad \displaystyle \tilde{q}^{i}_{nL} \sim \left( 1,
3, 1, 2, \frac{1}{3} \right) , \\[12pt] & \displaystyle u^{i}_{nR}
\sim \left( 3, 1, 1, 1, \frac{4}{3} \right) , & \qquad
\displaystyle \tilde{u}^{i}_{nR} \sim \left( 1, 3, 1, 1,
\frac{4}{3} \right) , \\[12pt] & \displaystyle d^{i}_{nR} \sim
\left( 3, 1, 1, 1, - \frac{2}{3} \right) , & \qquad \displaystyle
\tilde{d}^{i}_{nR} \sim \left( 1, 3, 1, 1, - \frac{2}{3} \right) ,
\\[12pt] & \ell_{nL} \sim (1, 1, 1, 2, -1) \; , & \qquad \tilde{\ell}_{nL}
\sim (1, 1, 2, 1, -1) \; , \\[12pt] & \nu_{nR} \sim (1, 1, 1, 1, 0) \; ,
& \qquad \tilde{\nu}_{nR} \sim (1, 1, 1, 1, 0) \; , \\[12pt] & e_{nR}
\sim (1, 1, 1, 1, -2) \; , & \qquad \tilde{e}_{nR} \sim (1, 1, 1,
1, -2) \; ,
\end{array}
\label{fermions}
\end{eqnarray}
where $i$ denotes the color degree of freedom, $n$ refers to the
three generations, and the numbers in brackets describe the
gauge-group transformation properties.

The hidden prequarks and their exotic partners, here denoted by
hats, have the following transformation qualities under the gauge
group:
\begin{eqnarray}
\begin{array}{rll}
& \hat{q}^{i}_{nL} \sim (3, 1, 2, 1, - 1) \; , & \qquad
\hat{\tilde{q}}^{i}_{nL} \sim (1, 3, 1, 2, - 1) \; ,  \\[12pt]
& \hat{u}^{i}_{nR} \sim (3, 1, 1, 1, 0) \; , & \qquad
\hat{\tilde{u}}^{i}_{nR} \sim (1, 3, 1, 1, 0) \; ,  \\[12pt] &
\hat{d}^{i}_{nR} \sim (3, 1, 1, 1, - 2) \; , & \qquad
\hat{\tilde{d}}^{i}_{nR} \sim (1, 3, 1, 1, - 2) \; .
\end{array}
\end{eqnarray}
In this scenario of prequarks, leptons, and exotic partners, the
gauge anomalies produced by the integer hypercharge of ordinary
and exotic prequarks are cancelled by incorporating local
counterterms with Chern--Simons configurations of gauge fields, as
it is done for the SM gauge group $\mbox{SU(3)}_{q} \times
\mbox{SU(2)}_{q\ell} \times \mbox{U(1)}_{Y}$.\cite{EAM21,EAM22}

Thus, on the one hand, the electroweak part of the bare Lagrangian
remains invariant under the extended $\hat{P}$-presymmetry
transformation, where ordinary (exotic) prequark multiplets of a
given color are exchanged with ordinary (exotic) lepton multiplets
according to
\begin{eqnarray}
\begin{array}{rcl}
& \hat{q}^{i}_{nL} \leftrightarrow \ell_{nL} \; , \qquad
\hat{u}^{i}_{nR} \leftrightarrow \nu_{nR} \; , \qquad
\hat{d}^{i}_{nR} \leftrightarrow e_{nR} \; , & \\[12pt] &
\hat{\tilde{q}}^{i}_{nL} \leftrightarrow \tilde{\ell}_{nL} \; ,
\qquad \hat{\tilde{u}}^{i}_{nR} \leftrightarrow \tilde{\nu}_{nR}
\; , \qquad \hat{\tilde{d}}^{i}_{nR} \leftrightarrow
\tilde{e}_{nR} \; , &
\end{array}
\label{Presymm}
\end{eqnarray}
and $W^{a}_{q} \leftrightarrow W^{a}_{\tilde{q}}$ for the gauge
bosons of $\mbox{SU(2)}_{q\tilde{\ell}}$ and
$\mbox{SU(2)}_{\tilde{q}\ell}$, respectively.  This discrete
symmetry requires that their gauge coupling constants be equal.

On the other hand, there is still invariance under the electroweak
$\hat{\tilde{P}}$-presymmetry transformation, where ordinary
(exotic) prequark multiplets of a given color are exchanged with
exotic (ordinary) leptons:
\begin{eqnarray}
\begin{array}{rcl}
& \hat{q}^{i}_{nL} \leftrightarrow \tilde{\ell}_{nL} \; , \qquad
\hat{u}^{i}_{nR} \leftrightarrow \tilde{\nu}_{nR} \; , \qquad
\hat{d}^{i}_{nR} \leftrightarrow \tilde{e}_{nR} \; , & \\[12pt]
& \hat{\tilde{q}}^{i}_{nL} \leftrightarrow \ell_{nL} \; , \qquad
\hat{\tilde{u}}^{i}_{nR} \leftrightarrow \nu_{nR} \; , \qquad
\hat{\tilde{d}}^{i}_{nR} \leftrightarrow e_{nR} \; , &
\end{array}
\label{exoticpresym}
\end{eqnarray}
keeping all the gauge fields unchanged. This discrete symmetry is
equivalent to the electroweak presymmetry at the level of the SM.

\begin{figure}[t]
\begin{picture}(300,70)
\put(90,18){$\tilde{\ell}_{nL}$}
\put(140,18){$\hat{\tilde{q}}^{i}_{nL}$} \put(117,18){$\hat{P}$}
\put(120,17){\vector(1,0){10}} \put(120,17){\vector(-1,0){10}}
\put(90,59){$\hat{q}^{i}_{nL}$} \put(140,59){$\ell_{nL}$}
\put(117,58){$\hat{P}$} \put(120,57){\vector(1,0){10}}
\put(120,57){\vector(-1,0){10}}
\put(87,35){$\hat{\tilde{P}}$} \put(95,40){\vector(0,1){10}}
\put(95,40){\vector(0,-1){10}} \put(147,35){$\hat{\tilde{P}}$}
\put(145,40){\vector(0,1){10}} \put(145,40){\vector(0,-1){10}}
\put(120,40){\vector(1,1){10}} \put(120,40){\vector(-1,-1){10}}
\put(120,40){\vector(-1,1){10}} \put(120,40){\vector(1,-1){10}}
\put(124,36){$\tilde{P}$} \put(115,0){(a)}
\put(225,18){$\hat{\tilde{\ell}}_{nL}$}
\put(275,18){$\tilde{q}^{i}_{nL}$} \put(252,18){$\hat{P}$}
\put(255,17){\vector(1,0){10}} \put(255,17){\vector(-1,0){10}}
\put(225,59){$q^{i}_{nL}$} \put(275,59){$\hat{\ell}_{nL}$}
\put(252,58){$\hat{P}$} \put(255,57){\vector(1,0){10}}
\put(255,57){\vector(-1,0){10}}
\put(221,35){$\hat{\tilde{P}}$} \put(229,40){\vector(0,1){10}}
\put(229,40){\vector(0,-1){10}} \put(282,35){$\hat{\tilde{P}}$}
\put(280,40){\vector(0,1){10}} \put(280,40){\vector(0,-1){10}}
\put(255,40){\vector(1,1){10}} \put(255,40){\vector(-1,-1){10}}
\put(255,40){\vector(-1,1){10}} \put(255,40){\vector(1,-1){10}}
\put(259,36){$\tilde{P}$} \put(250,0){(b)}
\end{picture}
\caption{Electroweak presymmetric transformations of fermion
doublets in the scenario of (a)~prequarks and (b) preleptons.}
\label{figure1}
\end{figure}
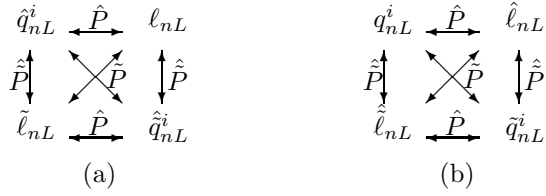

As shown in Fig.~\ref{figure1}(a), the $Z_2$ symmetries $\hat{P}$
and $\hat{\tilde{P}}$ introduce the exotic $\tilde{P}$-symmetry
defined by $\tilde{P} = \hat{\tilde{P}} \hat{P} = \hat{P}
\hat{\tilde{P}}$, under which ordinary and exotic prequarks and
leptons are transformed as follows:
\begin{eqnarray}
\begin{array}{rcl}
& \hat{q}^{i}_{nL} \leftrightarrow \hat{\tilde{q}}^{i}_{nL} \; ,
\qquad \hat{u}^{i}_{nR} \leftrightarrow \hat{\tilde{u}}^{i}_{nR}
\; , \qquad \hat{d}^{i}_{nR} \leftrightarrow
\hat{\tilde{d}}^{i}_{nR} \; ,
& \\[12pt] & \ell_{nL} \leftrightarrow \tilde{\ell}_{nL} \; , \qquad
\nu_{nR} \leftrightarrow \tilde{\nu}_{nR} \; , \qquad e_{nR}
\leftrightarrow \tilde{e}_{nR} \; , &
\end{array}
\label{exoticsym}
\end{eqnarray}
and $W^{a}_{q} \leftrightarrow W^{a}_{\tilde{q}}$ for the gauge
bosons, with equal gauge couplings.  Moreover, now the whole bare
Lagrangian of electroweak and strong interactions is invariant
under the exotic $\tilde{P}$-symmetry if we include the
transformation $G^{b}_{q} \leftrightarrow G^{b}_{\tilde{q}}$ for
the gluons of $\mbox{SU(3)}_{q}$ and $\mbox{SU(3)}_{\tilde{q}}$,
and make equal their couplings. In this manner, presymmetry
becomes extended from matter to forces, as demanded.

Ordinary and exotic quarks of fractional charge are generated from
ordinary and exotic prequarks of integer charge through universal
fractional charge shifts $\Delta Y = 4/3$ induced by integer
topological charges.\cite{EAM1}$^{\mbox{--}}$\cite{EAM22} In this
mechanism, $\hat{P}$ and $\hat{\tilde{P}}$ presymmetries are
broken. The exotic $\tilde{P}$-symmetry, however, remains exact.
This shows that it can be seen as a manifestation of a remaining
presymmetry in the sense of Ekstein\cite{Ekstein1,Ekstein2} and
its extension from matter to forces. Its spontaneous breaking
occurs because of the gauge symmetry breaking,\cite{EAM3} which
does not involve charges themselves.

Note that exotic partners are required to have a physical residual
presymmetry. If no extension of the SM with Dirac neutrinos is
considered, presymmetry remains hidden with no survival of
presymmetry relations between observable quarks and
leptons.\cite{EAM1} In the same way, the symmetric duplication of
the SM proposed in Ref.~\refcite{EAM3} may appear somewhat
contrived if there is no connection with a residual presymmetry.

\section{Presymmetric Extension of the SM with Prelepton
Partners}

The hidden electroweak presymmetry operates in two forms: between
prequarks and leptons, and between quarks and preleptons. In the
scenario of hidden preleptons and exotic partners, their
assignments to the gauge group $\mbox{SU(3)}_{q} \times
\mbox{SU(3)}_{\tilde{q}} \times \mbox{SU(2)}_{q\tilde{\ell}}
\times \mbox{SU(2)}_{\tilde{q}\ell} \times \mbox{U(1)}_{Y}$ are
\begin{eqnarray}
\begin{array}{rll}
& \displaystyle \hat{\ell}_{nL} \sim (1, 1, 1, 2, \frac{1}{3}) \;
, & \qquad \hat{\tilde{\ell}}_{nL} \sim (1, 1, 2, 1, \frac{1}{3})
\; , \\[12pt] & \displaystyle \hat{\nu}_{nR} \sim (1, 1, 1, 1, \frac{4}{3})
\; , & \qquad \hat{\tilde{\nu}}_{nR} \sim (1, 1, 1, 1,
\frac{4}{3}) \; , \\[12pt] & \displaystyle \hat{e}_{nR} \sim
(1, 1, 1, 1, - \frac{2}{3}) \; , & \qquad \hat{\tilde{e}}_{nR}
\sim (1, 1, 1, 1, - \frac{2}{3}) \; .
\end{array}
\end{eqnarray}

The preleptonic form of the $\hat{P}$-presymmetry transformation
in Eq.~(\ref{Presymm}) is set up as
\begin{eqnarray}
\begin{array}{rcl}
& q^{i}_{nL} \leftrightarrow \hat{\ell}_{nL} \; , \qquad
u^{i}_{nR} \leftrightarrow \hat{\nu}_{nR} \; , \qquad d^{i}_{nR}
\leftrightarrow
\hat{e}_{nR} \; , & \\[12pt] & \tilde{q}^{i}_{nL} \leftrightarrow
\hat{\tilde{\ell}}_{nL} \; , \qquad \tilde{u}^{i}_{nR}
\leftrightarrow \hat{\tilde{\nu}}_{nR} \; , \qquad
\tilde{d}^{i}_{nR} \leftrightarrow \hat{\tilde{e}}_{nR} \; , &
\end{array}
\end{eqnarray}
and $W^{a}_{q} \leftrightarrow W^{a}_{\tilde{q}}$.

In place of the $\hat{\tilde{P}}$-presymmetry in
Eq.~(\ref{exoticpresym}), we have
\begin{eqnarray}
\begin{array}{rcl}
& q^{i}_{nL} \leftrightarrow \hat{\tilde{\ell}}_{nL} \; , \qquad
u^{i}_{nR} \leftrightarrow \hat{\tilde{\nu}}_{nR} \; , \qquad
d^{i}_{nR} \leftrightarrow \hat{\tilde{e}}_{nR} \; , & \\[12pt]
& \tilde{q}^{i}_{nL} \leftrightarrow \hat{\ell}_{nL} \; , \qquad
\tilde{u}^{i}_{nR} \leftrightarrow \hat{\nu}_{nR} \; , \qquad
\tilde{d}^{i}_{nR} \leftrightarrow \hat{e}_{nR} \; , &
\end{array}
\end{eqnarray}
keeping the gauge fields unchanged.

Instead of the exotic $\tilde{P}$-symmetry transformation in
Eq.~(\ref{exoticsym}) we now obtain, as illustrated in
Fig.~\ref{figure1}(b),
\begin{eqnarray}
\begin{array}{rcl}
& q^{i}_{nL} \leftrightarrow \tilde{q}^{i}_{nL} \; , \qquad
u^{i}_{nR} \leftrightarrow \tilde{u}^{i}_{nR} \; , \qquad
d^{i}_{nR}
\leftrightarrow \tilde{d}^{i}_{nR} \; , & \\[12pt] & \hat{\ell}_{nL}
\leftrightarrow \hat{\tilde{\ell}}_{nL} \; , \qquad \hat{\nu}_{nR}
\leftrightarrow \hat{\tilde{\nu}}_{nR} \; , \qquad \hat{e}_{nR}
\leftrightarrow \hat{\tilde{e}}_{nR} \; , &
\end{array}
\end{eqnarray}
together with $W^{a}_{q} \leftrightarrow W^{a}_{\tilde{q}}$ and
$G^{b}_{q} \leftrightarrow G^{b}_{\tilde{q}}$.

Ordinary and exotic leptons of integer charges are generated from
ordinary and exotic preleptons of fractional charge in a way
similar to that described for quarks from prequarks.\cite{EAM1}
The exotic $\tilde{P}$-symmetry is again explained from a residual
presymmetry and its extension to forces, as posed in the
Introduction.

\section{Phenomenological Implications}

The exotic particles with $\tilde{P}$-symmetry have a
phenomenology similar to that of the partner particles in
supersymmetric models with $R$-parity, little Higgs models with
$T$-parity, and universal extra dimension models with $KK$-parity.
Since no partner particles have already been detected, all of
these models require symmetry breaking mechanisms in order to give
them a mass heavier than that of the SM particles. There exists a
large amount of work concerning their implications. For a number
of reasons, in particular the elegant solution to the hierarchy
problem, the supersymmetric partners appear as the leading
candidates to be the expected new particles.\cite{SS2}

Some features of our model, however, can make simple its
discrimination from the others. A spontaneous breaking of the
expanded gauge symmetry and the discrete exotic symmetry
$\mbox{SU(3)}_{q} \times \mbox{SU(3)}_{\tilde{q}} \times
\mbox{SU(2)}_{q\tilde{\ell}} \times \mbox{SU(2)}_{\tilde{q}\ell}
\times \mbox{U(1)}_{Y} \times \tilde{P}$ via the Higgs mechanism
has been discussed in Ref.~\refcite{EAM3}. It is a renormalizable
extension of the SM that includes a Higgs bidoublet and a
symmetric duplication of the SM Higgs doublet which produce the
breakdown at tree level and make mass of the new particles
different from known ones, and heavy enough as to have evaded
observation in experiments performed up to now. Although
constraints on masses of exotic partners are placed by cosmology
and precision electroweak experiments, no fine tuning at the
electroweak scale is required. It has been shown that three
generations of relatively heavy extra quarks and leptons can fit
the data in a two-Higgs-doublet scenario,\cite{He} with resulting
fermion masses greater than about 100 GeV and smaller than about 1
TeV. Such Higgs doublets are required to implement presymmetry in
the Higgs sector. The pressing mass limit for the lightest Higgs
boson is alleviated and the naturalness scale of the electroweak
model is ameliorated because the doubling of the SM Higgs doublet
can defer the fine-tuning problem to a higher scale, as argued in
Ref.~\refcite{Gunion}. Regarding the new exotic weak bosons, these
do not have the universality of the interactions of the standard
weak bosons, so that lower bounds of a few TeV can be set for
their masses.\cite{EAM3} And the new exotic gluons, which only
bind exotic quarks into exotic hadrons, are massless and have
asymptotically free couplings, just like the properties of usual
gluons in ordinary hadrons. The finding of a duplication of the
spectrum of the SM that keep spins and be nondegenerated and
natural enough around the TeV scale would be a strong support for
the model.

Nevertheless, to stabilize the hierarchy beyond this scale, even
new physics will be required. On the one hand, this would imply to
embed the extended gauge model into a supersymmetric model, or a
little Higgs model, or an extra dimension model, with the
corresponding proliferation of elementary gauge, scalar, and
fermionic particles. The alternative possibility of interpreting
the Higgs scalars as bound states of the extra heavy fermions, as
in technicolor models,\cite{Chivukula} appears as an attractive
minimalist approach. On the other hand, one can imagine a scenario
where the GUT scale is quite close to the Planck scale,
eliminating the hierarchy problem.\cite{Shapo} This would be an
interesting idea to pursue if no signal of supersymmetric
particles is found.

\section{Conclusions}

We have extended presymmetry from fermions to bosons and
consequently promoted the rather simple expansion of the SM with
Dirac neutrinos to the symmetric model $\mbox{SU(3)}_{q} \times
\mbox{SU(3)}_{\tilde{q}} \times \mbox{SU(2)}_{q\tilde{\ell}}
\times \mbox{SU(2)}_{\tilde{q}\ell} \times \mbox{U(1)}_{Y} \times
\tilde{P}$ of separate color and weak gauge groups for quarks,
leptons, and fermionic exotic partners. We have related the exotic
symmetry $\tilde{P}$, which exchanges the new particles and the SM
particles, with a residual presymmetry in the sense of Ekstein and
its extension from matter to forces. Constraints from high
precision experiments provide only restrictions on the mass of the
new particles. The upper bounds below 1~TeV for fermion partners
raise expectations of their direct detection. No fine-tuning
should be required at the electroweak scale.

In order to go further, we mention that presymmetry and exotic
symmetry apply to the forces of the SM implies duplication of
$\mbox{U(1)}_{Y}$ within the gauge group $G_{q \tilde{\ell}}
\times G_{\tilde{q} \ell} \times \tilde{P}$ with $G = \mbox{SU(3)}
\times \mbox{SU(2)} \times \mbox{U(1)}$. However, under a full
duplication of the SM, a residual presymmetry can also be realized
through $G_{q \ell} \times G_{\tilde{q} \tilde{\ell}} \times
\tilde{P}$, where now all SM particles are neutral with respect to
the hidden gauge group $G_{\tilde{q} \tilde{\ell}}$ and
$\tilde{P}$ is the hidden symmetry exchanging ordinary and hidden
partners; results on this alternative scenario will be reported
elsewhere. Hidden sectors have been invoked frequently in
extensions of the SM; for recent works see Ref.~\refcite{Cassel}.
This should be followed by a GUT for each $G$ to have
$(\mbox{GUT})_{q \tilde{\ell}} \times (\mbox{GUT})_{\tilde{q}
\ell} \times \tilde{P}$ in the exotic case and $(\mbox{GUT})_{q
\ell} \times (\mbox{GUT})_{\tilde{q} \tilde{\ell}} \times
\tilde{P}$ in the more standard hidden possibility, unifying
standard and exotic or hidden partners and also strong and
electroweak interactions. We note that a separation of GUT for
ordinary quarks and leptons assures a proton stability. GUT
predicts transitions between exotic and ordinary quarks and
leptons leading to the decay of heavy exotic matter into ordinary
matter, or transitions from heavy hidden matter to lighter hidden
matter. From a cosmological point of view, this is a mandatory
condition not to conflict with evidences on the absence of exotic
or hidden baryons. To stabilize the hierarchy up to the GUT scale
our model has to be embedded into, for example, a supersymmetric
model, or eliminate the problem by joining the GUT and Planck
scales. We also note that the well-known cosmological domain wall
problems associated with the spontaneous breakdown of the discrete
symmetry $\tilde{P}$ may be solved via scattering of primordial
black holes,\cite{SFS} with no additional constraints on the
model. Finally, we remark that the separation of electroweak and
strong gauge interactions for ordinary and exotic or hidden
partners opens the question of whether that somehow holds for
gravity. Its answer would urge to know the gravitational
properties of exotic or hidden matter.

\section*{Acknowledgments}

We thank J. Gamboa and L. Vergara for helpful comments. This work
was partially supported by the Departamento de Investigaciones
Cient\'{\i}ficas y Tecnol\'ogicas, Universidad de Santiago de
Chile, Usach.

\end{document}